\documentclass[a4paper,twoside]{article}

\usepackage{epsfig}
\usepackage{subcaption}
\usepackage{calc}
\usepackage{amssymb}
\usepackage{amstext}
\usepackage{amsmath}
\usepackage{amsthm}
\usepackage{multicol}
\usepackage{pslatex}
\usepackage{apalike}
\usepackage[bottom]{footmisc}
\usepackage{graphicx}
\usepackage{algorithm2e} 
\usepackage{amsfonts}
\usepackage{diagbox}
\usepackage{textcomp}
\usepackage{xcolor}
\usepackage{caption}
\usepackage{subcaption}
\usepackage{url}
\usepackage{multirow}
\usepackage[export]{adjustbox}
\usepackage[outdir=./]{epstopdf}
\usepackage{SCITEPRESS}     

\begin{document}

\title{A Robust Adaptive Workload Orchestration in Pure Edge Computing}

\author{\authorname{Zahra Safavifar\sup{1}, Charafeddine Mechalikh\sup{2} and Fatemeh Golpayegani\sup{1}}
\affiliation{\sup{1}University College Dublin, School of Computer Science, Ireland}
\affiliation{\sup{2}Laboratory of Artificial Intelligence and Information Technology (LINATI), University Kasdi Merbah, Ouargla, Algeria}
\email{zahra.safavifar@ucdconnect.ie, mechalikh.charafeddine@univ-ouargla.dz, fatemeh.golpayegani@ucd.ie}
}

\keywords{Workload orchestration, Reinforcement Learning, Pure Edge Computing, Adaptive model, Robust model.}

\abstract{Pure Edge computing (PEC) aims to bring cloud applications and services to the edge of the network to support the growing user demand for time-sensitive applications and data-driven computing. However, mobility and limited computational capacity of edge devices pose challenges in supporting some urgent and computationally intensive tasks with strict response time demands. If the execution results of these tasks exceed the deadline, they become worthless and can cause severe safety issues. Therefore, it is essential to ensure that edge nodes complete as many latency-sensitive tasks as possible.
\\In this paper, we propose a Robust Adaptive Workload Orchestration (R-AdWOrch) model to minimize deadline misses and data loss by using priority definition and a reallocation strategy. The results show that R-AdWOrch can minimize deadline misses of urgent tasks while minimizing the data loss of lower priority tasks under all conditions.}

\onecolumn \maketitle \normalsize \setcounter{footnote}{0} \vfill

\section{\uppercase{Introduction}}
\label{sec:introduction}

The rapid increase in actively connected Internet of Things (IoT) devices and their applications create massive data at the network's edge. A recent report estimates that connected devices will reach 38 billion by 2025 \cite{ref_url1}. The affordability and expansion of these devices have given rise to many IoT applications, including but not limited to connected vehicles, smart cities, healthcare, security, surveillance, and traffic monitoring. "Data Age" estimates that 175 zettabytes of data will be generated every year by 2025 of which 30\% will need real-time processing \cite{rao2021data}. While the cloud has huge resources, it suffers from increased latency, inconsistent connectivity, and a lack of real-time responsiveness. As a response, a new paradigm of Edge Computing has been introduced which involves the deployment of servers (or small-scale data centers) near the edge of the network. Increasing the number of such servers by increasing the data produced at the edge is not a solution. We need to shift toward Pure Edge Computing (PEC) to minimize the use of high-capacity servers and take advantage of currently available resources at the edge devices. However, edge devices mobility and their limited resources and energy may lead to the situation that available edge devices in the area cannot support all user's demands and increase failure. 

IoT applications can involve private data, require minimal latency, or produce a considerable volume of data while requiring fast execution. The failure to meet a critical deadline can lead to fatalities and significant losses \cite{uddin2021cloud,dai2019scheduling,khan2016survey}. This highlights the importance of resource management and scheduling in edge computing where a mix of requests with various Service Level Agreements (SLA) and deadlines should be managed with resource-constraint devices. The literature proposes different approaches to accommodate tasks with varying time sensitivity in Edge Computing. However, some dynamicity of the real-world environment is ignored for the sake of simplifying the problem. These simplifications include homogeneity of computing nodes and tasks  \cite{dai2019scheduling,sharif2022priority}, neglecting bandwidth fluctuation and its impact on delay \cite{sharif2022priority,fadahunsi2021edge,dai2019scheduling}, and disregarding the mobility of devices \cite{sharif2022priority,dai2019scheduling,xu2020computing,lee2021sdn,fadahunsi2021edge,azizi2022deadline}. Moreover, some of them assume simultaneous batch arrival of resource requests or formulate the problem as an offline scheduling problem by assuming a prior knowledge of task arrival times \cite{dai2019scheduling,sharif2022priority}. 
Fault tolerance is another concern in current distributed platforms where mission-critical IoT applications are involved, and the target edge node cannot function for any reason. Given the real-time nature of IoT applications, not handling such failures can cause a disaster. 

Currently, edge computing hosts a variety of IoT applications of which healthcare is attracting the attention of both researchers and industry players. Edge computing is being used to improve services in the healthcare domain. Since data in these platforms is related to human health safety, it can be very sensitive to latency and network dynamics (e.g., changes in available bandwidth) \cite{wang2017healthedge}. Pure Edge Computing, by taking advantage of available nearby devices as a computational resource, supports low latency services, which makes it a suitable option for deploying healthcare platforms. 

This paper applies workload orchestration in the healthcare platforms of a nursing home, where elderly people with varying health conditions are taken care of by monitoring and analyzing their health signals. This platform is facilitated by Pure Edge Computing where there are heterogeneous mobile and stationery edge devices such as smartphones, computers, wearables, sensors, etc., and a private Edge server in the area. As the generated tasks vary in urgency and priority, various Hard-Real-Time (HRT) tasks such as monitoring vital signals of bedridden patients, Soft-Real-Time (SRT) tasks such as processing camera data, and Non-Real-Time (NRT) tasks such as historical data gathering are generated and must be handled. 
Since healthcare applications are highly sensitive, and because the PEC environment is highly unpredictable and dynamic, a robust model is needed that can handle the workload in different conditions, including emergencies. It should be able to handle an environment that is sparse with a few devices, or when there are high loads of devices and requests. 

This paper proposes a Robust Adaptive Workload Orchestration (R-AdWOrch) that is designed based on AdWOrch \cite{safavifar2021adaptive}, which uses Reinforcement Learning for workload orchestration. The R-AdWOrch is able to function in various random dynamics in the PEC. The goal is to minimize the missing deadline of SRT tasks while meeting nearly all HRT tasks in any condition, specifically in sparse or crowded areas, by using a priority definition and a reallocation strategy. Our contributions to this paper are as follows:
\begin{itemize}
    \item Designing a reallocation strategy that prevents task failure when there are no available resources or the resource cannot execute the assigned task for any reason.  
    \item Prioritizing execution of HRT tasks over other types. 
    \item Reshaping the reward function to minimize the delay for HRT and SRT tasks.
\end{itemize}

The remainder of this paper is organized as follows, the PEC, Real-Time system, and AdWOrch are over-viewed in Section \ref{sec:Background}. The problem and environment characteristics are described in Section \ref{sec:ProblemStatement}. Task offloading process using R-AdWOrch is presented in Section \ref{sec:ModelDesign}. Section \ref{sec:PerformanceEvaluation} evaluates the proposed model through extensive simulations. Finally, Section \ref{sec:conclusion} concludes by giving an outline of the future directions of this work.

\section{\uppercase{Background}}
\label{sec:Background}

\begin{figure*}[]
  \centering
  \includegraphics[width=0.8 \textwidth]{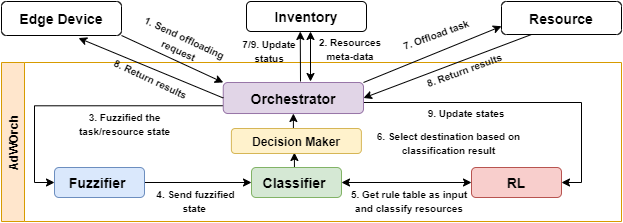}
  \caption{AdWOrch Components and task offloading flow.}
  \label{fig:component diagram}
\end{figure*} 

\subsection{Pure Edge Computing}
\label{sec:Pure-Edge-Computing}
Pure Edge Computing takes place at the bottom layer of the network architecture. It aims to improve the system's performance by taking advantage of available resources at the edge of the network.
Based on a study conducted by Carnegie Mellon University, computing at the extreme edge is more energy and latency efficient than computing in the distant cloud for certain applications \cite{drolia2013case}. 
However, PEC is a highly decentralized and dynamic network of heterogeneous and resource-constrained devices that can be mobile or stationary, which produce unpredictable and fluctuating workloads. IoT applications vary and are requested at different time sensitivity. Therefore, such an environment requires an orchestrator that can adapt to different situations and dynamics.

\subsection{Real-Time and Non-Real-Time Systems}
\label{sec:Real-Time and Non-Real-Time systems}

A real-time system is one whose basic specification and design correctness arguments must include its ability to meet its timing constraints \cite{kim1995predictability}. In contrast, non-real-time systems have no strict deadlines, and their tasks can be failed and repeated if they are not complete. Three levels of time sensitivity systems and tasks are defined:

\textbf{Hard-Real-Time (HRT)} tasks have firm deadlines. Failure to meet these deadlines can cause severe damage to the system or its environment, including injury or even death. Aviation control, fire alarm system, and some subsystem of the health care system are instances of hard real-time systems. 

\textbf{Soft-Real-Time (SRT)} tasks have soft/flexible deadlines in which delays or failure in execution lead to downgrading the Quality of Service. Video streaming and online gaming are two examples of soft real-time systems.

\textbf{Non-Real-Time (NRT)} tasks do not have restricted deadlines, and the system can tolerate the failure of such tasks. Data collection for some applications, such as weather prediction and road maintenance, can be categorized as non-real-time systems. 

Real-world smart platforms usually contain a mixture of different types of tasks which should be managed based on priority.

\subsection{AdWOrch}
\label{AdWOrch}
The Adaptive Workload Orchestration (AdWOrch) model is developed based on the Fuzzy Decision Tree (FDT) algorithm proposed in \cite{Mechalikh2020}.The base task orchestration algorithm is complemented by a new resource selection strategy and reward function. 

Figure \ref{fig:component diagram} shows the AdWOrch process and components.  First, the edge device sends an offloading request to the central orchestrator. The orchestrator gets all nearby resources' characteristics from the inventory and creates a state observation from the tasks' attributes and resources' characteristics and sends it to the fuzzifier. After the fuzzification phase, the resources are classified to estimate how much each resource is reliable for offloading the task using a decision tree. A decision tree is created using the Q-table that arises from the RL component. Then the decision-maker selects a nearby device with sufficient reliability. After offloading the tasks to the destination, resource status in the inventory is updated simultaneously. The result is sent to the orchestrator, which in turn sends it to the corresponding device. In addition, when the result is returned, the device's resource status in the inventory is updated again and the task execution result is sent back to the RL component. Finally, the RL component updates the Q-table based on the received results by the reward function.  

\section{\uppercase{Problem Statement}}
\label{sec:ProblemStatement}

This study aims to propose a robust and adaptive workload orchestration model to minimize missed soft real-time deadlines while meeting almost all hard real-time deadlines and minimizing data loss due to mobility or any other reasons to fail (i.e., insufficient resources, no available resources, and dead devices) in a PEC environment. 

The PEC environment is where resource-constrained devices should accommodate a mix of various tasks with different urgency levels and SLA. Also, devices are heterogeneous, and both end devices and resources can be stationary or mobile. Mobile devices are able to enter and leave the area or move within it effortlessly. Battery-powered devices differ in their remaining battery level and locations.

\subsection{Parameter Modelling}
\label{sec:ParameterModelling}

A centralized powerful RL-based orchestrator entity allocates or reallocates tasks to the available resources by considering the type of the tasks (i.e., HRT, SRT, NRT)  to minimize the delay and failure in the systems.

We proposed two tiers architecture in which edge devices and sensors at the first layer generate the tasks and are utilized as the primary computational resources. While for managing emergent conditions in which devices at the edge cannot satisfy the requirements of the user request, the Edge servers at the second layer will facilitate the required resources for HRT and SRT tasks. In this paper, the state, action, and reward for the RL algorithm are defined as follows:
\begin{itemize}
    \item \textbf{State representation} 
    In AdWOrch, the state is represented by two feature sets, tasks' attributes and resources' characteristics at timestamp $t$. A task's attributes include latency, size, generator mobility, and task type. Resources' characteristics include CPU utilization, CPU MIPS, mobility behavior, and remaining battery. To improve the learning process, instead of CPU utilization, we consider the current load on the device, $d_l$, which is calculated by the queue length of tasks waiting for execution on the device, $d_{ql}$, divided by the number of device CPU cores, $d_{cc}$. See Equation. \ref{eq:current load}.
    \begin{equation}\label{eq:current load}
                  d_l = d_{ql} / d_{cc} 
    \end{equation}
   Instead of device capacity that was the MIPS size and task size, the expected execution time, $e_t$,is considered which is calculated by the task size, $t_s$, divided by the MIPS size, $d_m$. See Equation. \ref{eq:expected execution time}.
    \begin{equation}\label{eq:expected execution time}
                  e_t = t_s / d_m 
    \end{equation}
    \item \textbf{Action space}
    The action space $A=\{0, 1\}$ is the set of all possible resource selection strategies (i.e., indicating a specific resource that has been selected for the task). 
    \item \textbf{Reward}
    The immediate reward the orchestrator gets from the environment at a time step $t$ after taking action $a$ is the success or failure status of the task, which is represented as 1 or 0, respectively. 
\end{itemize}

\section{\uppercase{Model Design}}
\label{sec:ModelDesign}

\begin{figure}
  \centering
  \includegraphics[width=7cm]{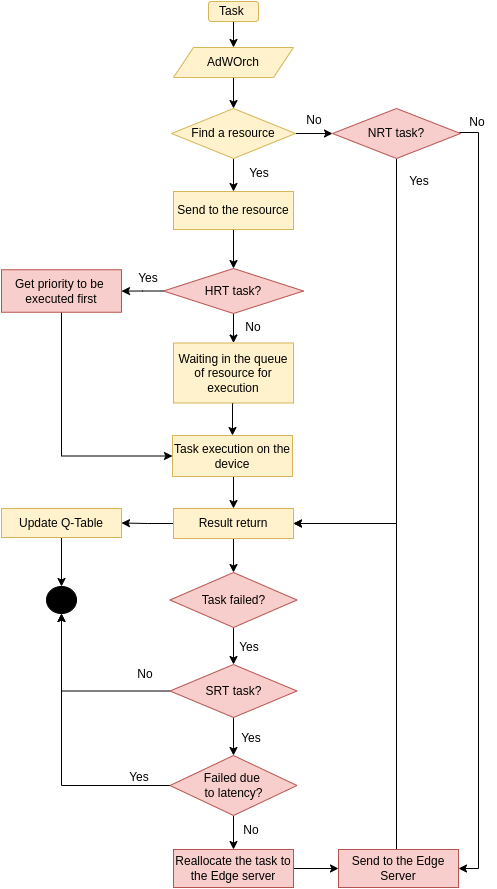}
  \caption{Task offloading process by R-AdWOrch}
  \label{fig:R-AdWOrch}
\end{figure}

\begin{figure*}[]
  \centering
  \includegraphics[width=0.7\textwidth]{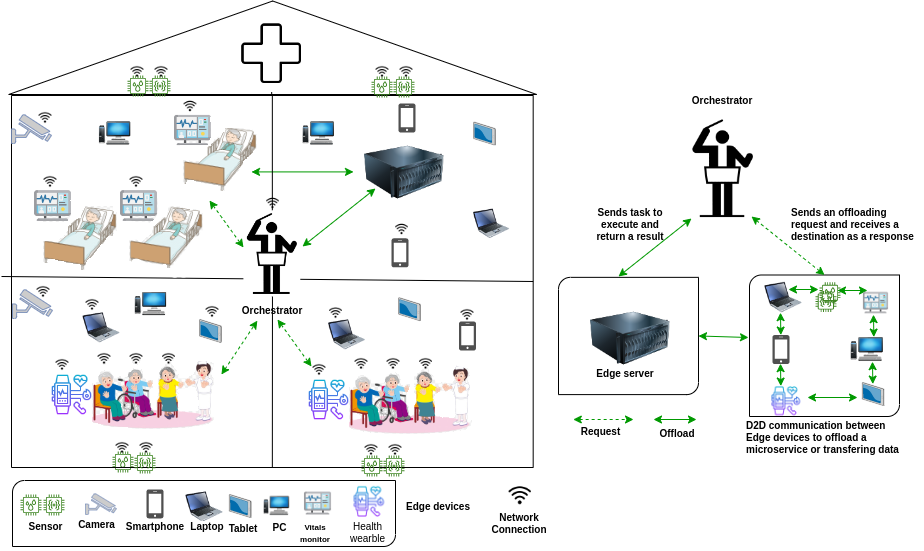}
  \caption{Health care platform in a PEC environment}
  \label{fig:HealthcareInPEC}
\end{figure*} 

A Robust Adaptive Workload Orchestration (R-AdWOrch) model is proposed based on AdWOrch model \cite{safavifar2021adaptive}. It aims to achieve robustness along with adaptability in random and dynamic PEC environments which can accommodate resource constraints in crowded or sparse areas.
R-AdWOrch introduces a central orchestrator that employs an RL-based algorithm that aims to minimize the missing deadline for SRT tasks and meet nearly all HRT tasks. Hence, it should handle different types of task failure. There are five types of failure in the system:
\begin{itemize}
    \item \textbf{Failure due to missing the deadline:} a task cannot be completed before its deadline. 
    \item \textbf{Failure due to mobility:} mobile devices can join or leave the area effortlessly and not be able to access one another to return the result during the offloading process. 
    \item \textbf{Failure due to incompatible hardware/software:} various devices in the area have different hardware and software scopes. Since tasks have different requirements in terms of hardware and software, sometimes the allocated device scope might not be compatible with the task requirements. It causes the task to fail due to insufficient resources. 
    \item \textbf{Failure due to no available resources:} In the PEC environment, a device might be isolated and cannot find nearby devices to offload a task. So, the task fails due to no available resources. 
    \item \textbf{Failure due to a dead device:} When a battery-operated device runs out of energy and cannot executes offloaded tasks. 
\end{itemize}

Figure \ref{fig:R-AdWOrch} demonstrates how R-AdWOrch manages different types of tasks (i.e., HRT, SRT, NRT) to minimize task failure for the above failure reasons. The R-AdWOrch has added red components to AdWOrch. When R-AdWOrch receives a task, it tries to find a proper resource from surrounding devices using the AdWOrch orchestrator. If a proper resource is found, the task will be sent to the resource. In the case of no accessible resources, there are two possibilities: when the task is an NRT task, it will fail immediately due to the "no available resources". Otherwise, if it is an HRT or SRT task, it will be sent to the Edge server directly to prevent failure due to "No available resources". The R-AdWOrch ignores all NRT tasks in this step for two reasons. i) When a disaster happens, there are a few resources and limited bandwidths which should be mainly utilized for executing Real-Time tasks. ii) If the Edge server and bandwidth are overloaded with NRT tasks, it might lead to long response time and therefore the failure of the HRT task. 

When an offloading destination is selected, the task is offloaded to that device for execution. If it is an HRT task, it will be  a high priority to execute before all SRT and NRT tasks. Hence, it decreases the chance of HRT tasks failure due to "missing deadline". Since a preemptive system is considered, the NRT task execution is paused and returned back to the queue to allow arrived HRT task to be executed immediately. NRT and SRT tasks should wait in the device queue for execution. After execution, the result returns to calculate the new q-value and update the q-table (see \ref{sec:RewardFunction}). When a failed result is returned, if the task is an SRT and failed for other reasons than latency, such as incompatible hardware/software, mobility, and dead device, it will be reallocated to the Edge server to be completed before the deadline. Finally, the result returns after execution at the Edge server.  

Also, to avoid failure due to the dead device, all SRT and NRT tasks will be reallocated when the battery level drops below a given threshold. Current HRT tasks will still be executed but the orchestrator will prevent offloading future tasks to this device (see Algorithm 1). 

\subsection{Reward Function}
\label{sec:RewardFunction}
Our goal is to minimize the SRT task failure rate while meeting nearly all HRT deadlines. The AdWorch \cite{safavifar2021adaptive} defined a delayed penalty in the reward function to minimize the failure due to missing a deadline. This work adds more penalties for HRT and SRT task failure as is shown in Equ. \ref{eq:Reward function}.
\begin{equation}\label{eq:Reward function}
              R(s,a) = R_s - (w * D_p)  
\end{equation}

\vspace{-0.3cm}\begin{equation}\label{eq:delay penalty}
              D_p = T_d / T_l 
\end{equation}

$R_s$ is the task's execution result, which is in the set of $\{0, 1\}$to indicate success or failure. The delay penalty $D_p$ is calculated by dividing the delay time $T_d$ by the task latency time $T_l$ (see Equation \ref{eq:delay penalty}). The $D_p$ value can exceed $1$ but this rarely occurs. $w$ is the weight of the penalty which for the HRT task is $3$, for the SRT task is $1.5$ and for the NRT tasks is $1$.

\begin{algorithm}[hbt!]
\label{alg:reallocationAlg}
\footnotesize
\caption{Tasks reallocation for low battery edge devices } \label{alg:alg2}
\While{running}{
        \If{offloading request $(task, device)$}{
            \If{remaining energy$(device) <$ threshold}{
                \eIf{$task \neq HRT$}{
                    reallocate($task$,INSUFFICIENT\_POWER)\; /*reallocate due to insufficient power*/
                }{ 
                    addToExecutionQueue($t$)\;
                }
            }
        }
    }
\end{algorithm}

\begin{table*}[hbt!]
\centering
\caption{Types of edge devices}\label{tab1}
\begin{tabular}{|l|c|c|c|c|c|}
\hline
Types &  Laptops & Smartphones & Gateways & Stationary sensors & Mobile sensors\\
\hline
Generate tasks & No & Yes & No & Yes & Yes\\
Ratio (\%) & 11 & 18 & 11 & 28 & 32\\
Mobility & No & Yes & No & No & Yes\\
Speed (m/s) & - & 1.4 & - & - & 1.4\\
Battery-powered & Yes & Yes & No & No & Yes\\
Battery-capacity(Wh) & 56.2 & 18.75 & - & - & - \\
Idle energy consumption (W) & 1.7 & 0.2 & 3.8 & - & -\\
Max energy consumption (W) & 23.6 & 5 & 5.5 & - & -\\
CPU (GIPS) & 110 & 25 & 16 & - & -\\
CPU cores & 8 & 8 & 4 & - & -\\
\hline
\end{tabular} 
\end{table*}

\section{\uppercase{Performance Evaluation}}
\label{sec:PerformanceEvaluation}

\subsection{Simulation Settings}
\label{sec:SimulationSettings}

To evaluate the performance of R-AdWOrch in a pure edge environment we use PureEdgeSim \cite{mechalikh2020pureedgesim}.
We simulate a smart nursing home where senior citizens with different health conditions live. Some are bedridden, while others are in better condition and can move around and do everyday tasks. Various healthcare sensors collect different vitals and send the data for analysis and diagnosis. The urgency of data differs; for example, the vitals of a critically ill patient has a very high priority compared with those of a healthy individual. In addition to health sensors, other devices such as cameras are used to record people's apparent activities, and some sensors collect historical data from the equipment and environment. The characteristics of these devices are summarized in Table \ref{tab1}.

An emergency in the mentioned smart nursing home can happen when a telecommunication incident happens. Half of the edge devices are disconnected from the main network while still having access to the internal network. Also, when for any reason the number of people is increased, the system faces a high workload. (see Table \ref{tb:Scenarios profile}). A private Edge server is in the area, and devices can communicate with it via the main network. This server will have a computing capacity of 400 GIPS (Giga Instruction Per Second). Monitoring vital signs of critically ill patients are HRT tasks, while analyzing vitals from healthy individuals and processing camera data are SRT tasks. The tasks of gathering historical data are considered as NRT (see Table \ref{tab2}). There is a central orchestrator that all devices can access via the internal network and accesses the main network and the Edge server. 

To evaluate the performance of the proposed model, the results are compared to AdWOrch \cite{safavifar2021adaptive}. For each of the explained scenarios, a simulation of 30 minutes is run 5 times. 

\subsection{Evaluation Metrics\\}
We define two evaluation metrics. The \textbf{task success rate} for each task type is calculated by the number of successful tasks divided by the total number of generated tasks. The \textbf{average delay time} for each type of task is calculated by dividing the total delay time by the number of generated tasks. The total delay time is calculated by summing up the total network time, total execution time, and total waiting time. Furthermore, we assessed how effective the \textbf{reallocation} strategy is by calculating how many tasks are reallocated due to low battery power, mobility, and incompatible resources. 

\begin{table}[h]
\caption{The types of applications}\label{tab2}
\begin{center}
\begin{tabular}{|c|c|c|}
\hline
Application types & Latency(ms) & Size (MI)\\
\hline
HRT&15& 200\\
SRT&500 & 5000\\
NRT&30000& 10000\\ 
\hline
\end{tabular} 
\end{center}
\end{table}

\begin{table}[hbt!]
\caption{Scenarios profile}
\label{tb:Scenarios profile}
\begin{center}
\begin{tabular}{|c|c|c|c|c|}
\hline
\cline{1-4} 
\textbf{Number of devices} & \textbf{HRT $^{\mathrm{a}}$}& \textbf{SRT $^{\mathrm{a}}$}& \textbf{NRT $^{\mathrm{a}}$} \\
\hline
50&135&135&270 \\
100&135&135&270 \\
300&135&135&270\\
\hline
\multicolumn{4}{l}{$^{\mathrm{a}}$Requested task for each device per minute.}\\
\end{tabular}
\label{tab3}
\end{center}
\end{table}

\subsection{Results and Discussion}
\label{sec:ResultsAndDiscussion}
\begin{figure*}[hbt!]
 \centering 
  \captionsetup[sub]{font=small,labelfont={bf,sf}}
  \begin{subfigure}{0.33\textwidth}
  \includegraphics[width=\linewidth]{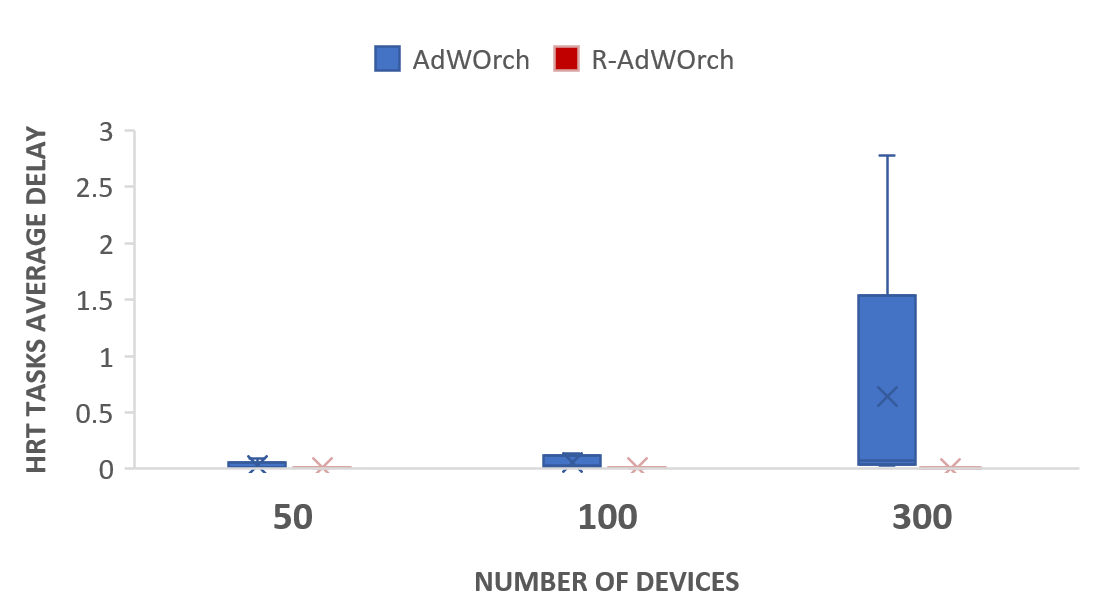}
  \caption{Average HRT delay}
  \label{fig:4}
\end{subfigure}\hfil 
\begin{subfigure}{0.33\textwidth}
  \includegraphics[width=\linewidth]{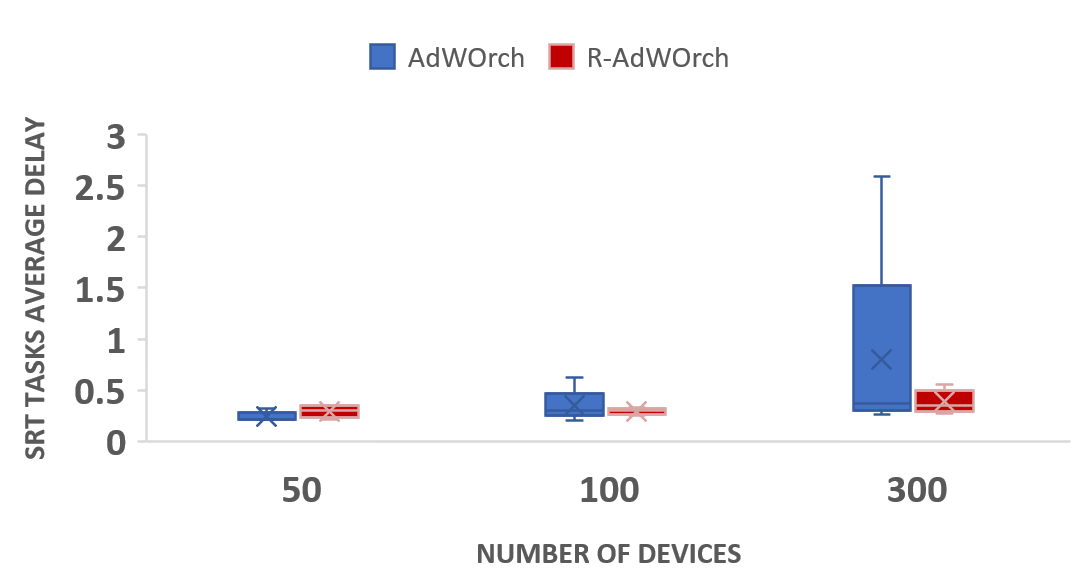}
  \caption{Average SRT delay}
  \label{fig:5}
\end{subfigure}\hfil 
\begin{subfigure}{0.33\textwidth}
  \includegraphics[width=\linewidth]{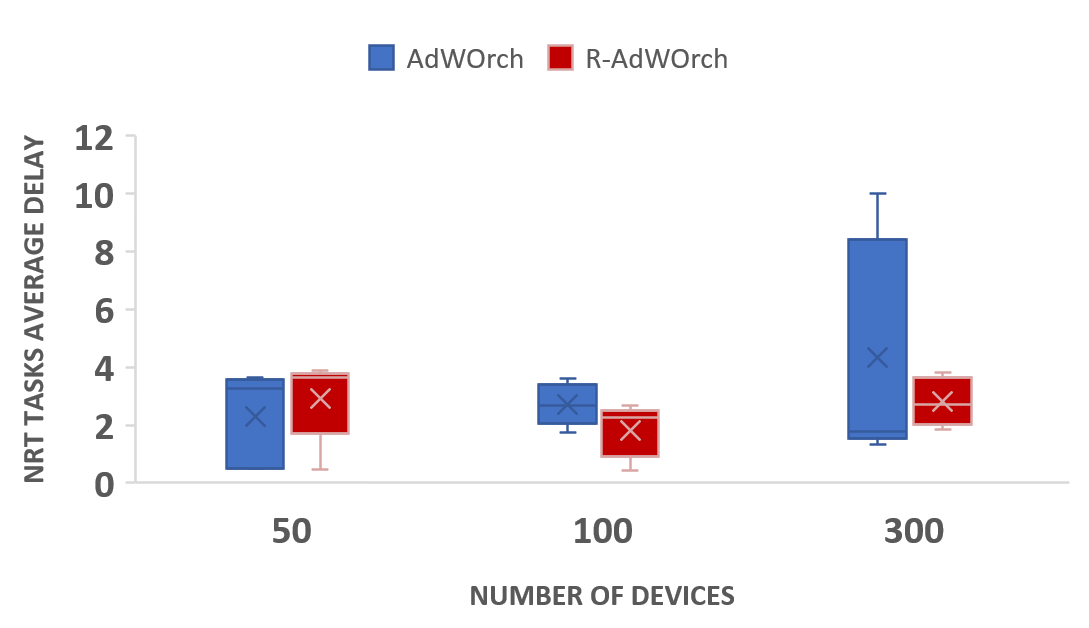}
  \caption{Average NRT delay}
  \label{fig:6}
\end{subfigure}
\medskip
\begin{subfigure}{0.33\textwidth}
  \includegraphics[width=\linewidth]{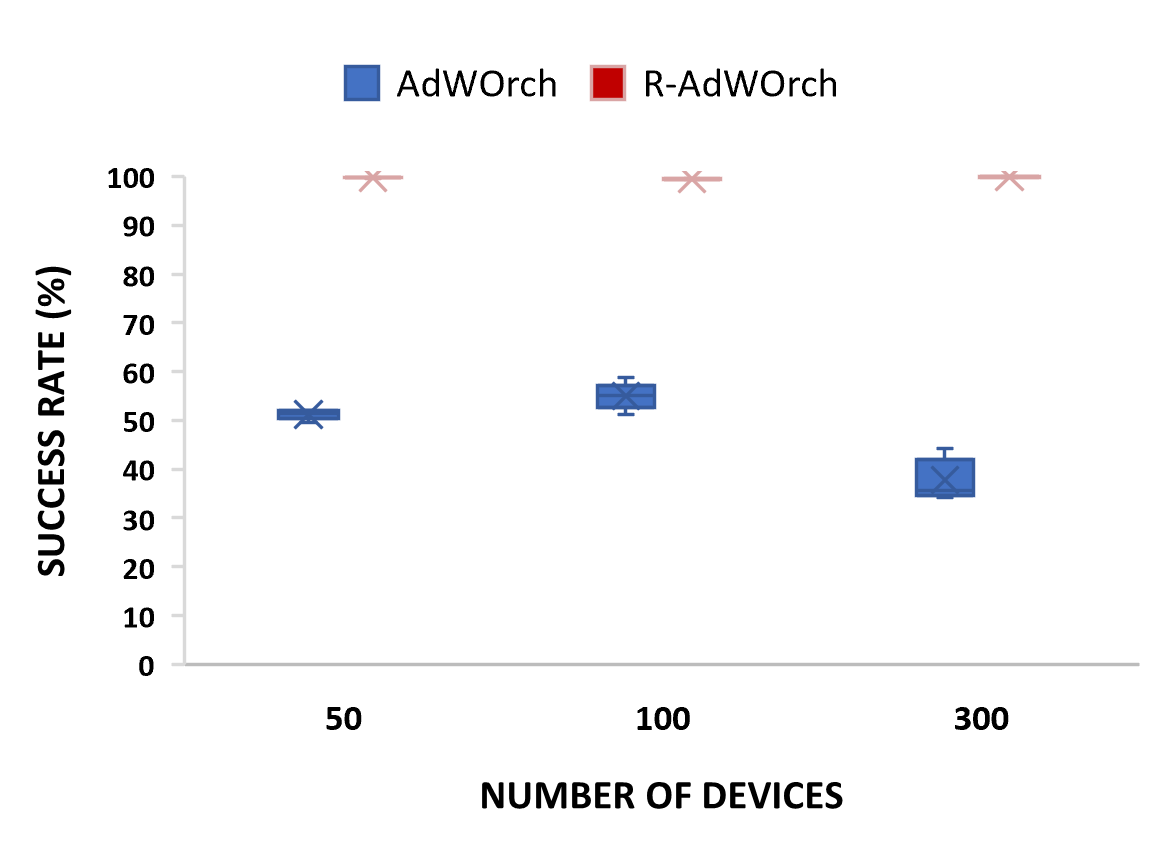}
  \caption{HRT success rate}
  \label{fig:1}
\end{subfigure}\hfil 
\begin{subfigure}{0.33\textwidth}
  \includegraphics[width=\linewidth]{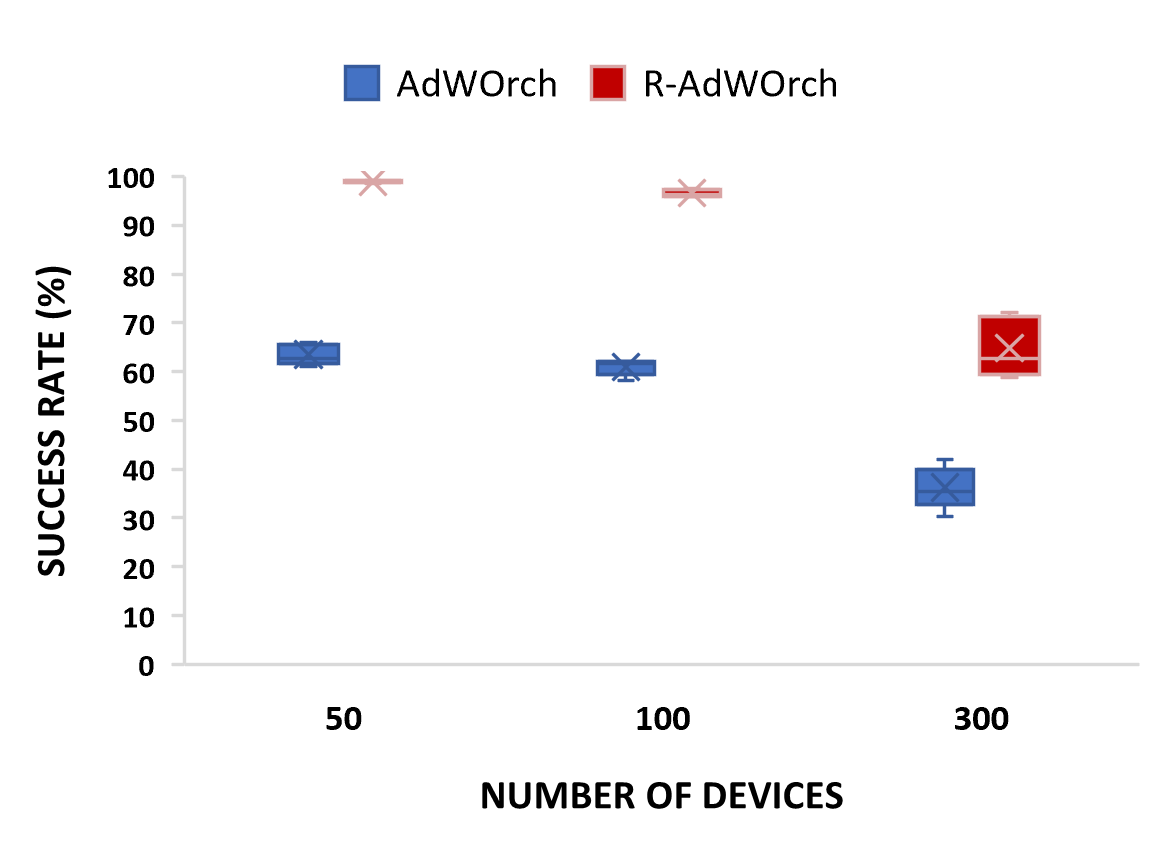}
  \caption{SRT success rate}
  \label{fig:2}
\end{subfigure}\hfil 
\begin{subfigure}{0.33\textwidth}
  \includegraphics[width=\linewidth]{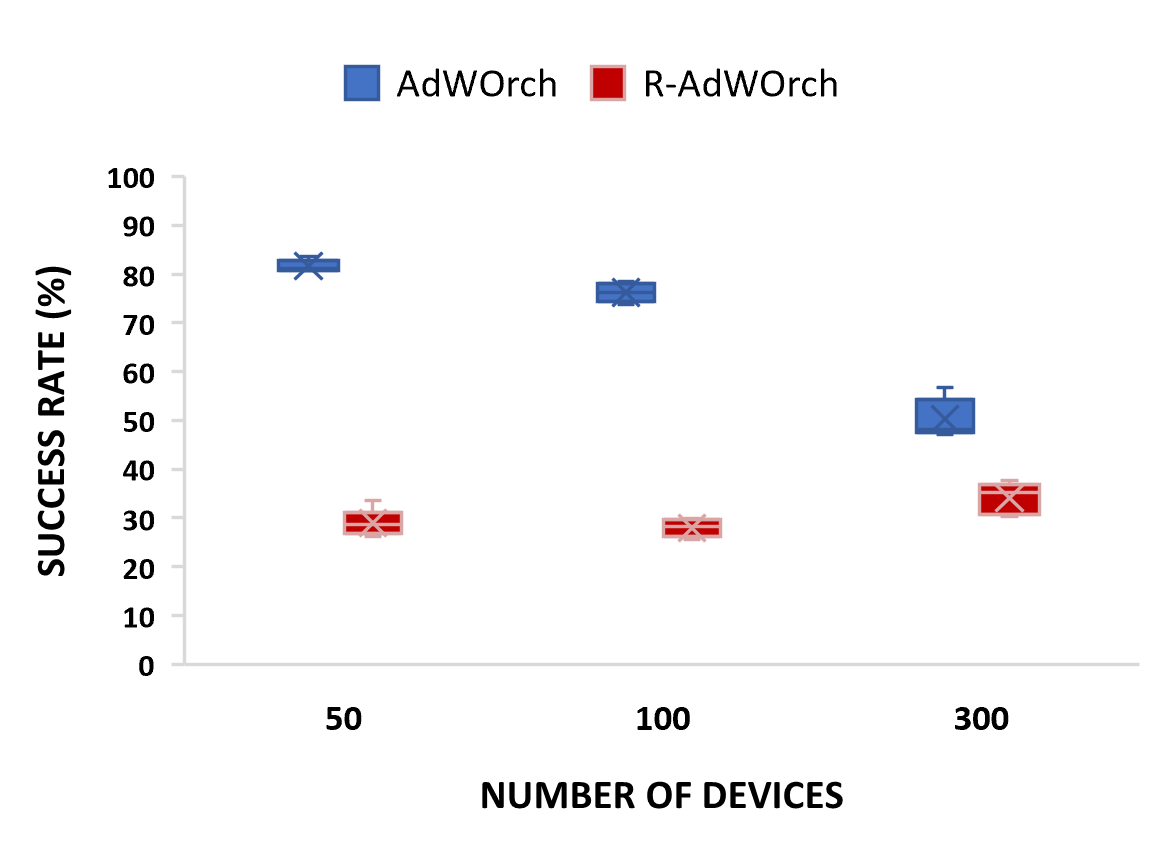}
  \caption{NRT success rate}
  \label{fig:3}
\end{subfigure}
\caption{The simulation results for different scenarios.}
\label{fig:results}
\end{figure*}

\textbf{Average Delay:}
Figure \ref{fig:results}(a-c) shows the average delay for each type of task under different workloads. As can be seen, R-AdWOrch significantly minimizes the delay of HRT tasks compared to the baseline when the density of devices is low (i.e., 50 devices) while providing comparable results in all other cases. One of the reasons for the reduction of the delay is the extra penalty that we defined in the reward function for HRT and SRT task failure. The RL component receives a heavily weighted penalty when an HRT task is not completed within the expected time interval. Moreover, HRT tasks are prioritized over non-HRT in two ways: i) the HRT tasks will add to the front of the queue; ii) when an HRT task arrives, if no free computational resource is available, executing NRT tasks are paused and resources are allocated to HRT tasks. These allow fast execution even in resource-constrained scenarios.

\textbf{Success rate:}
Figure \ref{fig:1} shows that R-AdWOrch increases the success rate of HRT tasks compared to the baseline. The reason is R-AdWOrch applies two levels of prioritization to prevent HRT tasks from failing: i) by reassigning non-HRT tasks to other devices when the remaining battery power is not enough for executing all its remaining tasks; ii) as mentioned in section \ref{sec:RewardFunction}, the delay penalty for the HRT tasks is multiplied by 3 which causes the algorithm to be more sensitive to HRT failures compared with SRT and NRT tasks. Moreover, as can be seen in Figure \ref{fig:2} R-AdWOrch success rate for SRT tasks is significantly higher than the baseline. The main reason is the allocation strategy that applies to SRT tasks when they fail due to reasons other than latency. By reallocating these tasks to the edge server, they can meet the deadline and succeed. Figure \ref{fig:3} shows the NRT tasks baseline outperforms the R-AdWOrch which is reasonable because i) in R-AdWOrch due to the nature of NRT tasks they have lower priority for execution; ii) they do not send to the edge server when no nearby edge devices are available for assigning and they will be failed.

\begin{figure}[htb!]
  \centering
  \includegraphics[width=0.4\textwidth]{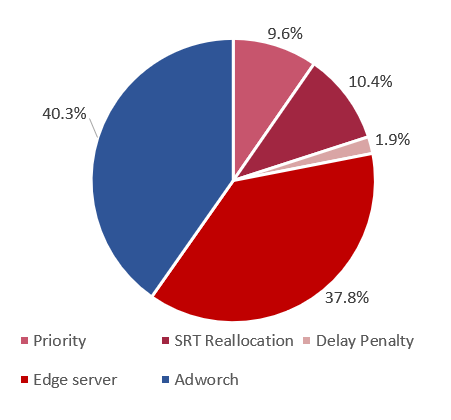}
  \caption{The contribution of R-AdWOrch methods in the success rate.}
  \label{fig:PieChartSuccess}
\end{figure}

Figure \ref{fig:PieChartSuccess} depicts the contribution of each of the proposed methods (i.e., priority, reallocation, SRT/HRT delay penalty, and edge server) in improving the task success rate. In emergency scenarios, AdWOrch accounts for only 40\% of the success rate while the other 60\% comes from R-AdWOrch improving methods.

\begin{figure}[htb!]
  \centering
  \includegraphics[width=0.4\textwidth]{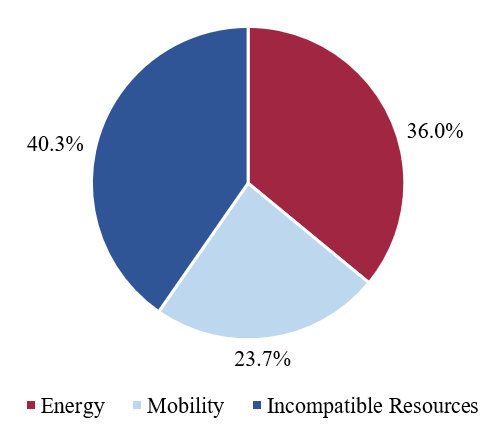}
  \caption{The contribution of reallocation reasons.}
  \label{fig:PieChartReallocations}
\end{figure}

\textbf{Reallocation:} Figure \ref{fig:PieChartReallocations} gives the number of reallocated SRT tasks and their reallocation reasons. As mentioned above, R-AdWOrch reallocates SRT tasks for two reasons: i) to avoid data loss when an SRT task is failed due to ``mobility'' or ``incompatible hardware/software''; ii) to prevent task failure resulting from a battery-powered device dead. 

\section{\uppercase{Conclusions}}
\label{sec:conclusion}

By considering the real-world characteristics of the PEC, explained above, this paper proposed a Robust Adaptive Workload Orchestration (R-AdWOrch) in a PEC environment that can function in different situations and circumstances. It aims to meet nearly all HRT tasks while minimizing the deadline missing of SRT tasks. We applied our model for a healthcare application and compared it with the baseline model (AdWOrch). The results show that R-AdWOrch outperforms AdWOrch in HRT and SRT success rates and decreases the delay time for these task types.

R-AdWOrch uses a centralized orchestrator which might be a single point of failure and is a risk for a robust model. Designing a distributed network of orchestrators is part of our future work. Moreover, the orchestration of dependent tasks is another challenge in this area. 

\section*{\uppercase{Acknowledgements}}

This publication has emanated from research supported in part by a grant from Science Foundation Ireland under Grant number 18/CRT/6183. For the purpose of Open Access, the author has applied a CC BY public copyright license to any Author Accepted Manuscript version arising from this submission.

\bibliographystyle{apalike}
{\small
\bibliography{maintext}}

\end{document}